\documentclass[11pt]{article}

\usepackage[T1]{fontenc}
\usepackage[utf8]{inputenc}
\usepackage{lmodern}
\usepackage[margin=1in]{geometry}
\usepackage{microtype}
\usepackage{graphicx}
\usepackage{float}
\usepackage{booktabs}
\usepackage{array}
\usepackage{multirow}
\usepackage{makecell}
\usepackage{xcolor}
\usepackage{colortbl}
\usepackage{enumitem}
\usepackage[most]{tcolorbox}
\usepackage[numbers,sort&compress]{natbib}
\usepackage[hidelinks]{hyperref}

\setlist{leftmargin=*,itemsep=2pt,topsep=2pt}

\definecolor{quadblue}{RGB}{230,240,255}
\definecolor{quadgreen}{RGB}{232,245,233}
\definecolor{quadyellow}{RGB}{255,248,225}
\definecolor{quadgray}{RGB}{242,242,242}

\newtcolorbox{contribbox}{
  colback=white,
  colframe=blue!65!black,
  boxrule=0.8pt,
  arc=2pt,
  left=6pt,
  right=6pt,
  top=4pt,
  bottom=4pt
}

\begin{document}

\title{Audio-Based Condition Monitoring for Agricultural Machinery: Stakeholder Needs, Adoption Barriers, and MVP Requirements}

\author{
Tashfain Ahmed\thanks{Tashfain Ahmed and Mohammadali Saffary contributed equally to this work.}
\quad Mohammadali Saffary\footnotemark[1]
\quad Kehinde Elelu\\
Duc Hung Luong \quad Joshua Siegel\\[0.5em]
\small DeepTech Lab, Department of Computer Science and Engineering\\
\small Michigan State University, East Lansing, Michigan, USA\\
\small \texttt{\{ahmedt11,saffarym,elulukeh,luongduc,jsiegel\}@msu.edu}}

\date{}

\maketitle

\begin{abstract}
Failures of agricultural machinery are costly when breakdowns occur during narrow work windows and are discovered too late for planned service. Audio-based condition monitoring offers a low-cost retrofit sensing path because operators already treat abnormal sound as an informal diagnostic cue. Whether such systems are adopted, however, depends on how well they fit farm workflows, repair constraints, and stakeholder decision-making. This paper reports a stakeholder discovery study of audio-based condition monitoring for agricultural machinery. We synthesize semi-structured interviews with farm operators, dairy operators, dealership and service personnel, manufacturers, integrators, and domain experts, together with a short survey. Across sources, downtime avoidance emerged as the dominant value driver, and failures were most commonly recognized through unusual noise and loss of functionality. Survey responses further indicate a gap in maintenance visibility and low penetration of predictive tools. These findings narrow the opportunity from generic ``AI diagnosis'' to low-friction early warning and evidence generation for maintenance planning. Modern tractors appear to be a weaker initial wedge because many are already instrumented and supported by mature service programs. Implements, attachments, hidden rotating components, small-to-medium operations, and continuous-use settings such as dairy present clearer need because failures in these contexts are harder to perceive, sensing coverage is lower, and downtime consequences are higher. We translate these findings into requirements for a minimum viable product (MVP) that prioritizes maintenance actionability over full automation: commodity audio capture, contextual annotation, asset-specific baselines, persistent-deviation detection, temporally stable alerts, and evidence artifacts that support operator inspection and dealer triage.
\end{abstract}


\section{Introduction}
This paper examines where audio-based condition monitoring creates value in agricultural machinery and what an initial system must provide to be adopted. Agricultural operations depend on machinery that must perform reliably during narrow and often unforgiving production windows. A breakdown during planting, harvest, feeding, or milking creates costs that extend beyond the failed component itself, including delayed work, idle labor, emergency service calls, rental needs, and downstream operational losses. In this setting, the timing of detection matters: a small mechanical change that could be inspected during planned maintenance becomes costly when it is noticed only after performance drops or the machine stops.

Audio-based condition monitoring offers a practical path toward earlier detection because sound is already part of how operators recognize machinery problems. Farmers and equipment operators often notice abnormal noise, vibration, squealing, grinding, or changes in machine rhythm before they can articulate a precise diagnosis. This cue, however, is informal, subjective, and difficult to document or compare over time. A low-friction audio system could formalize this existing practice by capturing short recordings, linking them to operating context, and tracking whether a perceived change persists.

The value of such a system depends on more than sensing feasibility or model accuracy. Agricultural machinery is heterogeneous, service access is uneven, and diagnostic coverage varies by asset type. Newer tractors are often supported by telematics, fault codes, planned maintenance programs, and dealer workflows, whereas implements, attachments, belts, bearings, pumps, and other moving subsystems are less instrumented and harder to observe during operation. Even when diagnostic data exist, operators may face proprietary tooling, technician shortages, long scheduling delays, and limited repair access. These constraints make adoption a question of fit: which assets have sufficient unmet monitoring need, and what output would change an operator's next maintenance decision?

We address this question through customer discovery involving crop and dairy operators, dealership and service personnel, manufacturers, integrators, and domain experts, complemented by a short survey. Rather than assuming that technical feasibility implies demand, we use stakeholder evidence to identify the operating contexts, equipment segments, adoption barriers, and workflow requirements that should shape an audio-based monitoring pipeline. In doing so, the study connects the priorities of agricultural technology users with the decisions made by technology designers.

\begin{figure}[htbp]
\centering
\includegraphics[width=0.7\textwidth]{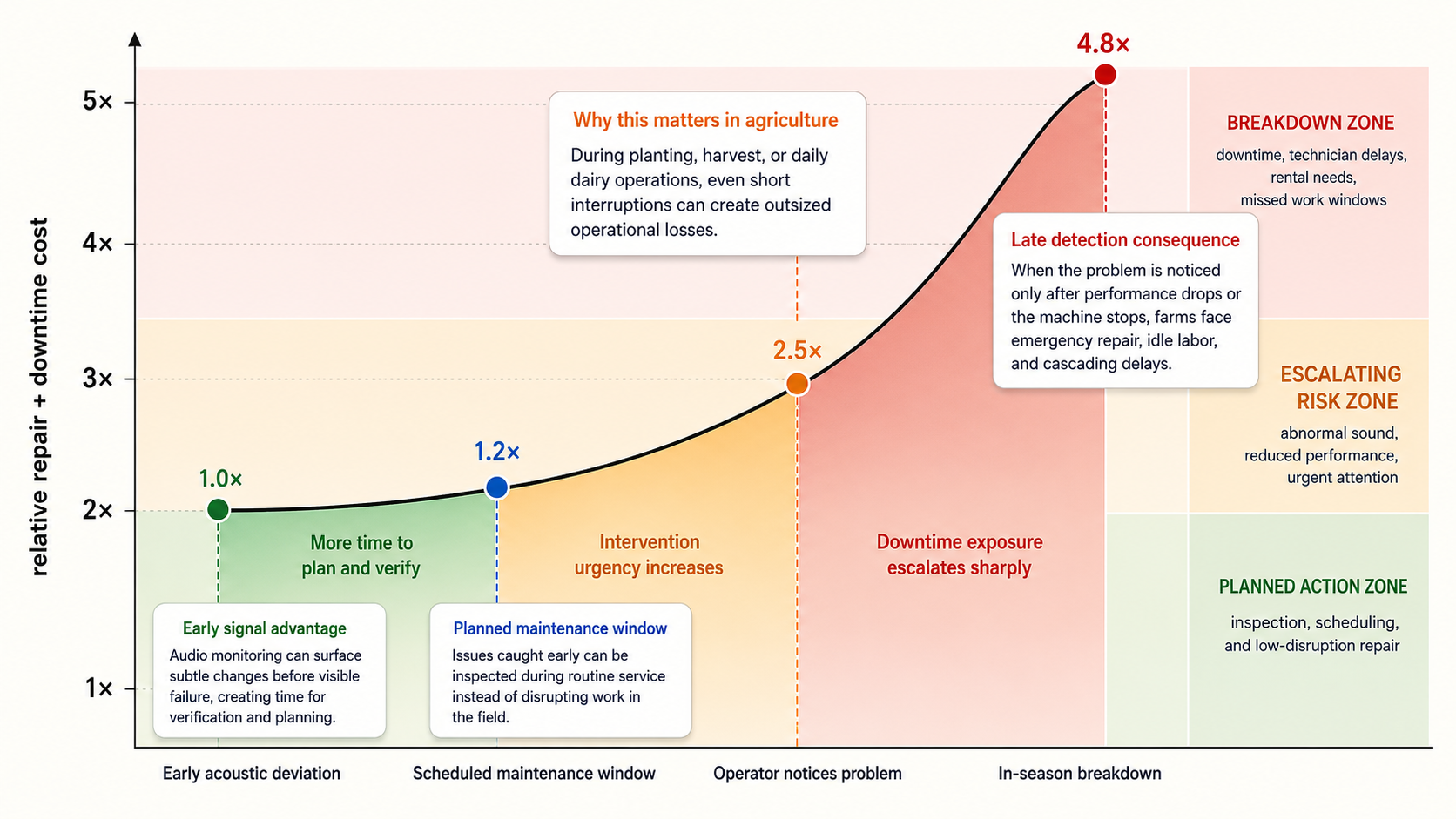}
\caption{Relative repair and downtime cost across detection stages. The cost of responding to a fault can rise nonlinearly as detection moves from early acoustic deviation to in-season breakdown. Early detection supports planned inspection and repair, whereas late detection compounds costs through downtime, emergency labor, expedited parts, and cascading damage. The three zones (Planned Action, Escalating Risk, and Breakdown) represent progressively shorter windows for intervention as an unaddressed fault becomes more severe~\cite{DOE2010OM,GitNux2026,Hanna2023,PIRG2023,IBM2020PdM}.}
\label{fig:stakeholder_chart}
\end{figure}

The findings indicate an initial market narrower than farm equipment in general. High-urgency contexts with limited monitoring coverage include implements and attachments, hidden rotating components, small-to-medium operations with limited service capacity, and continuous-use environments such as dairy. In these settings, failures are difficult to perceive, downtime is costly, and existing sensing or service coverage is often incomplete. Figure~\ref{fig:stakeholder_chart} illustrates the economic rationale for early detection: the cost of responding to a fault can rise nonlinearly as detection moves from an early acoustic deviation to an in-season breakdown.

\begin{contribbox}
\textbf{Contributions.} This paper contributes a stakeholder-grounded problem formulation for audio-based condition monitoring in agricultural machinery. Specifically, it provides:
(1) qualitative evidence on stakeholder needs, maintenance workflows, and downtime constraints;
(2) adoption-focused design implications identifying where audio monitoring is most likely to add value; and
(3) MVP requirements for a practical audio-monitoring pipeline, including low-friction capture, contextual annotation, asset-specific baselines, stable alerts, and shareable evidence for service triage.
\end{contribbox}

\section{Background}
This section situates the study within prior work on condition-based maintenance, digital agriculture adoption, and acoustic sensing, and identifies the gap the present study addresses. Predictive maintenance and condition-based maintenance (CBM) have long been studied as alternatives to purely schedule-based servicing \cite{Jardine2006CBM,Carvalho2019PdMReview}. The central premise is that degradation can often be detected before failure if systems observe useful indicators and support timely action \cite{Jardine2006CBM,Randall2011BearingTutorial}. Prior surveys emphasize that reliable decision support requires more than fault detection: systems must also handle limited sensing, uncertain labels, changing operating conditions, and lifecycle-cost constraints \cite{Jardine2006CBM,Carvalho2019PdMReview,Liu2018AIFaultDiagnosis}. Although much of this work originates in industrial environments with dedicated sensing infrastructure, agricultural machinery operates under more variable field conditions, with heterogeneous assets, changing loads, environmental noise, and uneven access to connectivity and service support \cite{Mattetti2022CANBUS,Bettucci2024WorkingStates}.

Adoption research in digital agriculture indicates that technical capability alone does not determine uptake. Digital and precision agriculture have expanded the data available to farms through guidance systems, yield monitors, variable-rate tools, telematics, and remote imagery \cite{Wolfert2017BigData,Klerkx2019DigitalAg,McFadden2023ERS}, yet adoption remains uneven across technologies and production contexts \cite{LowenbergDeBoer2019,Pierpaoli2013PAAdoption,McFadden2023ERS}. Prior work shows that agricultural technology adoption depends on perceived usefulness, ease of implementation, compatibility with existing workflows, and the availability of support infrastructure such as connectivity, training, and service networks \cite{McFadden2023ERS,LowenbergDeBoer2019,Thomas2023TAMReview,Pierpaoli2013PAAdoption,Klerkx2019DigitalAg}. These factors are especially consequential for equipment-monitoring cyber-physical systems (CPS), where a technically capable sensing system may still fail if it disrupts seasonal work, produces unclear outputs, or does not fit how operators and service personnel make maintenance decisions \cite{Rose2016DSSTools,Eastwood2019SocioEthical}.

This literature consistently frames adoption as a socio-technical problem involving economics, infrastructure, skills, reliability, and trust \cite{Klerkx2019DigitalAg,Eastwood2019SocioEthical}. Recent syntheses identify economic constraints, connectivity limitations, technical knowledge gaps, reliability concerns, and privacy or security concerns as recurring barriers \cite{Dibbern2024DriversBarriers,Hundal2023IoTBarriers,Eastwood2019SocioEthical}. Decision-support research similarly shows that tools often fail when outputs are not trusted, interpretable, or aligned with farmer decision cycles \cite{Rose2016DSSTools}. For condition monitoring, this implies that system outputs must be actionable: operators need to know whether a change is persistent, whether it warrants inspection, and what evidence can be shared with a dealer or technician \cite{Jardine2006CBM,Rose2016DSSTools,Henriquez2014AudioVibration}.

Maintenance and repair realities make this problem particularly acute in agriculture. Downtime during planting and harvest is costly because even short delays can cascade into yield, scheduling, and labor losses \cite{Parsons1981Downtime}. At the same time, modern agricultural equipment increasingly relies on software-enabled diagnostics and proprietary repair tools \cite{ORielly2021Deere,FTC2021NixingFix}. Policy and practitioner analyses argue that restricted diagnostic access increases repair cost and delay, reducing operator resilience \cite{ORielly2021Deere,FTC2021NixingFix}, and recent regulatory and legal actions have highlighted repair restrictions as a systemic issue for equipment owners who depend on timely service \cite{Reuters2025FTCDeere}. In this context, monitoring creates value by giving operators earlier evidence for inspection, scheduling, and service triage \cite{Jardine2006CBM,Rose2016DSSTools}.

From a technical perspective, there is growing interest in using on-machine data, including CAN-bus telemetry and machine learning, to infer equipment states and improve agricultural operations \cite{Bettucci2024WorkingStates,Mattetti2022CANBUS}. Audio is a complementary sensing modality because it is non-contact, low cost, and deployable with commodity microphones \cite{Mba2006AcousticEmission,Henriquez2014AudioVibration,Jombo2023AcousticMonitoring}. It can capture signatures associated with rotating components, belts, bearings, pumps, airflow, friction, and other mechanical processes \cite{Randall2011BearingTutorial,Mba2006AcousticEmission,Jombo2023AcousticMonitoring}. Field deployment, however, raises questions that model performance alone cannot answer: which failure modes matter most, which assets lack adequate monitoring, what false-alarm tradeoffs are acceptable, and what evidence is sufficient to change maintenance behavior \cite{Carvalho2019PdMReview,Liu2018AIFaultDiagnosis,Rose2016DSSTools}.

In summary, prior work establishes the potential of condition monitoring and the importance of adoption-aware design, but does not translate these findings into requirements for audio-based monitoring of agricultural machinery \cite{Jardine2006CBM,Carvalho2019PdMReview,Klerkx2019DigitalAg,Rose2016DSSTools,Henriquez2014AudioVibration}. The present study addresses this gap by using stakeholder perspectives to define a problem formulation that can ground subsequent CPS sensing and modeling work.

\section{Study Design}
We designed the study to connect stakeholder pain points with concrete requirements for an audio-based machinery monitoring MVP. The study focused on three questions: where monitoring would create operational value, which barriers would shape adoption, and which outputs would support real maintenance workflows.

We conducted semi-structured interviews and administered a short survey covering six areas: (i) critical equipment and recurring failure modes; (ii) how operators currently notice or diagnose problems; (iii) the relative importance of inefficiency detection versus impending-failure detection; (iv) existing maintenance practices and visibility gaps; (v) adoption constraints, including service access, training, proprietary tooling, warranty, and workflow burden; and (vi) stakeholder reactions to the plausibility of audio-based monitoring. Interviews were captured as transcripts or structured notes. 

\textbf{Participants.} We collected fifteen interview and conversation records (I1--I15) through purposive and snowball sampling, with initial contacts made via the MSU DeepTech Lab network, the Michigan MTRAC program, and direct outreach at the Michigan Ag Expo in Grand Rapids (December 2024). Formal interviews lasted approximately one hour; shorter expo conversations were captured as structured notes. Stakeholder roles included crop and dairy operators, dealership and service personnel, equipment manufacturers and integrators, and domain experts in field data collection and agricultural technology adoption (Table~\ref{tab:sources}). In addition, we collected ten survey responses (I16--I25) from farmers and researchers. Recruitment prioritized role diversity over statistical representativeness, and all participants are identified by anonymized role-based codes.

\begin{table}[htbp]
\centering
\caption{Anonymized study sources.}
\label{tab:sources}
\begin{tabular}{@{}ll@{}}
\toprule
ID & Stakeholder type \\
\midrule
I1 & Farm operator \\
I2 & Dairy operators \\
I3 & Manufacturer \\
I4 & Manufacturer \\
I5 & Manufacturer \\
I6 & Dealership and service manager \\
I7 & Dealership and service staff \\
I8 & Domain expert \\
I9 & Manufacturer\\
I10 & Manufacturer notes \\
I11--I15 & Mixed manufacturer/expert/farmer notes \\
I16--I25 & Survey responses from farmers and researchers \\
\bottomrule
\end{tabular}
\end{table}

\textbf{Analysis.} We analyzed the data through thematic synthesis in three passes. First, we extracted salient statements, operational pain points, and concrete examples from each transcript or note record. Second, we coded these excerpts inductively into five categories: value drivers, target assets and failure modes, maintenance workflows and gaps, adoption barriers, and MVP requirements. Codes emerged from the data rather than from a predetermined framework. Third, we validated themes by checking whether patterns reported by one stakeholder type (e.g., farm operators) held independently across others (e.g., dealers, manufacturers), treating cross-role convergence as a marker of robustness. Survey responses served as a final triangulation step to assess whether the qualitative patterns corresponded to reported priorities in the broader sample. Where a theme appeared in only one source or role, we note this explicitly in the results.

\section{Results}

This section presents the main patterns that emerged from the stakeholder discovery process. The results are organized thematically, moving from broad value drivers to specific adoption implications for audio-based condition monitoring. Table~\ref{tab:survey} summarizes the survey results that anchor the qualitative themes.

\begin{table}[htbp]
\centering
\caption{Survey summary. Respondents rated impending-failure identification as highly important, reported limited visibility into maintenance needs, showed low prior exposure to predictive tools, and most commonly recognized failures through noise or loss of functionality.}
\label{tab:survey}
\small
\begin{tabular}{p{0.68\columnwidth}p{0.22\columnwidth}}
\toprule
Metric & Value \\
\midrule
Mean importance: identifying inefficiencies (1--5) & 4.2 \\
Mean importance: identifying impending failures (1--5) & 4.5 \\
Not aware of current maintenance and repair needs & 60\% \\
Have not tried diagnostic and predictive tools & 90\% \\
Report noticing failures via noise & 60\% \\
Report noticing failures via loss of functionality & 50\% \\
\bottomrule
\end{tabular}
\end{table}

\subsection{Downtime, not efficiency, is the dominant value driver}
Across stakeholders, downtime avoidance emerged as the dominant value driver rather than marginal efficiency gains. Survey respondents rated the importance of identifying impending failures higher than the importance of identifying inefficiencies (means of 4.5/5 and 4.2/5, respectively), and interviewees consistently linked failures to operational time pressure.

Dairy-specific interviews emphasized that equipment often runs more than 12 hours per day and can be near-continuous in large operations (e.g., extended milking schedules), leaving little room for extended downtime. When a machine fails, farms may need to arrange backups or rentals and incur substantial unplanned costs. This context increases the value of early warning but also raises the cost of false alarms: an alert must be credible enough to justify interruption, inspection, or scheduling. This asymmetry becomes a central constraint in the requirements derived in Section~5.

\subsection{Operators already use sound, but informally}
A consistent finding is that operators already rely on acoustics as a primary diagnostic cue. In the survey, the most commonly reported detection mechanisms were noise (6/10) and loss of functionality (5/10). Several interviewees described noticing that ``something was off'' but lacking tools to localize or quantify the issue early enough to act before escalation.

\begin{quote}
\small ``By the time you realize it is too late, \ldots you got to have somebody come out \ldots like a computer and everything.'' (I1, farm operator)
\end{quote}

This existing listening practice provides the behavioral basis for the MVP. A monitoring system can make the practice measurable, repeatable, and reviewable by quantifying deviation from a baseline, suppressing isolated acoustic variation, and preserving the clip that triggered concern for later comparison or discussion with service personnel.

\subsection{Maintenance visibility remains limited despite interest in early warning}
A central quantitative result is a gap in equipment-health visibility. In the survey, 6/10 respondents (60\%) reported that they were \emph{not} aware of the current state of repair and maintenance needs of their equipment. Several responses indicated reliance on reactive repair (``when something fails'') or on limited internal capability (e.g., ``I'm not a mechanic'').

Penetration of predictive and diagnostic tools was also low: 9/10 respondents reported never having tried a diagnostic or predictive maintenance solution. The combination of high stated importance for early failure identification and low prior adoption suggests that an MVP must be simple, impose low burden, and demonstrate clear return on investment. Its outputs should indicate what to do next rather than emphasize algorithmic novelty.

\subsection{Repair constraints make evidence more valuable than labels}
The interview data indicate that minor issues frequently become major downtime because of constraints outside the operator's control. The most common constraints were (i) proprietary diagnostic access, (ii) technician shortages and long scheduling delays, and (iii) geographic mismatch between where a product is serviced and where it is used.

Proprietary diagnostics were described most vividly by operators discussing modern diesel subsystems and computerized controls, where faults can force reduced performance modes and require dealer software to clear.

\begin{figure}[h]
    \centering
    \includegraphics[width=1.0\linewidth]{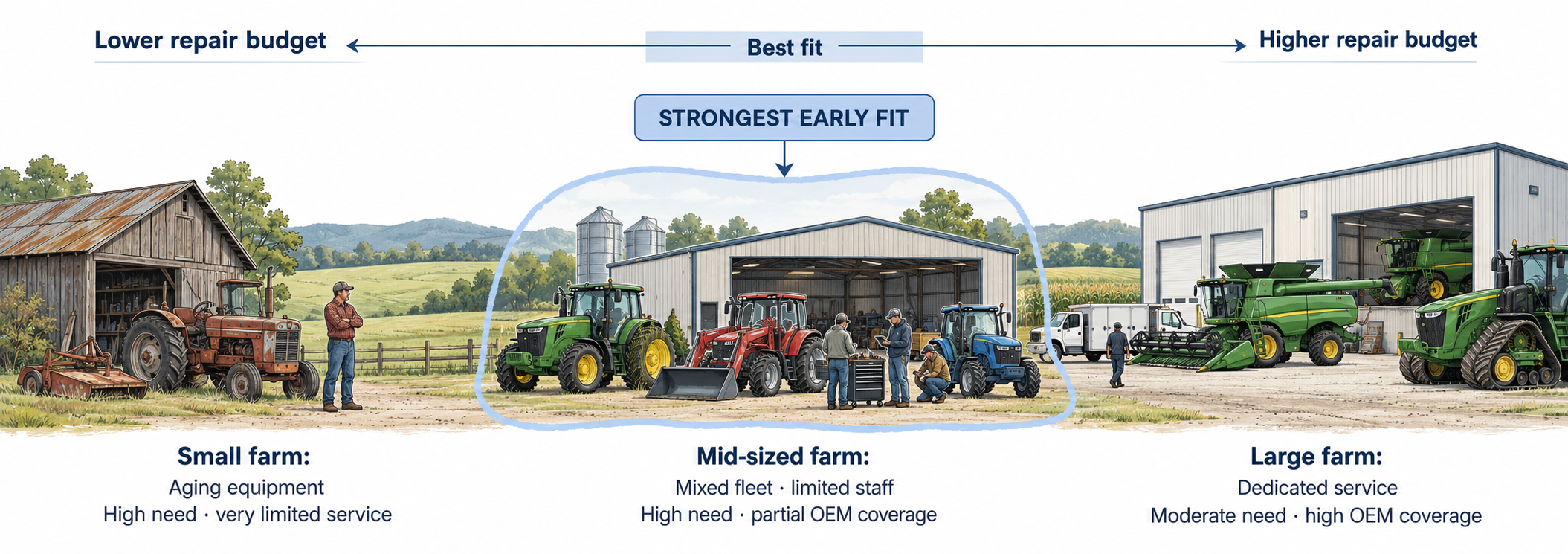}
    \caption{Qualitative fit assessment. Small and mid-sized operators appear to be a stronger early-adopter segment because they face higher downtime exposure and have less internal maintenance capacity than large farms.}
    \label{fig:operator_segment}
\end{figure}

\begin{quote}
\small ``You got to call [the dealer] \ldots It's all right-to-repair.'' (I1, farm operator)
\end{quote}

Several notes also indicate that diagnostic access can be expensive, training-intensive, and sometimes effectively limited to warranty contexts: one manufacturer noted that diagnostics are expensive and available ``only under warranty,'' and another reported that the software is hard to access and requires training. Technician shortages amplify these constraints, because even when a fault is understood, scheduling delays can mean missing a seasonal window.

Geography-driven service constraints appeared repeatedly in manufacturer conversations. Companies with strong local service in their home region reported difficulty supporting customers in distant states, particularly smaller and medium farms without dedicated on-site maintenance staff. Large farms often maintain in-house service capability, reducing their sensitivity to these delays, whereas smaller farms are more exposed. These constraints imply that a monitoring system's value lies less in naming a fault than in producing evidence that improves the service interaction.

\subsection{Segmentation: where monitoring creates value}

The results point to implements, attachments, and continuous-use systems outside mature original equipment manufacturer (OEM) diagnostic ecosystems as the highest-value segments. Figure~\ref{fig:segment_map} summarizes this segmentation along two dimensions: downtime urgency and existing sensing or service coverage.

\textbf{Modern tractors are a weaker initial wedge.} Dealership and service personnel expressed direct skepticism that acoustic monitoring would add value on modern tractors, noting that these machines already include extensive sensing, warranty-driven maintenance schedules, and planned service programs spanning annual maintenance, fluids, system checks, and mobile service visits. They described tractors as engineered to reliably complete critical seasonal windows.

\begin{quote}
\small ``No issues with tractors that an acoustic device could help identify.'' (I6, dealership)
\end{quote}

The same sources immediately identified clearer need in equipment pulled behind tractors: balers, bladed implements, and other multi-moving-part systems whose failures are not audible from inside modern quiet cabs and can escalate into safety incidents.

\begin{quote}
\small ``You cannot hear the bearing squeal \ldots the hay catches on fire.'' (I7, dealership)
\end{quote}

\textbf{Hidden subsystems are the acoustic opportunity.} Across manufacturer, integrator, and operator interviews, the most frequently cited failure candidates were bearings (including PTO-related bearings), belt tension changes, seal wear, and pump degradation. These are mechanically promising targets: bearing defects and belt tension shifts produce distinctive spectral and temporal signatures, and one manufacturer explicitly described wanting to distinguish tight, loose, and marginally functional belt regimes by sound (I5). These failures are difficult to perceive from within cab environments and often go undetected until downtime or a safety incident occurs. Early detection therefore has compounding value: wear in one component redistributes load and accelerates degradation elsewhere, so identifying deviations early interrupts a cascade rather than addressing a single fault.

\begin{figure}[h]
\centering
\scriptsize
\setlength{\tabcolsep}{4pt}
\renewcommand{\arraystretch}{1.12}
\begin{tabular}{|p{0.46\columnwidth}|p{0.46\columnwidth}|}
\hline
\cellcolor{quadgreen}
\begin{minipage}[t]{\linewidth}\vspace{4pt}
\centering
\textbf{High urgency + low coverage}\\[2pt]
(\textbf{Best early wedge})\\[4pt]
\begin{itemize}[leftmargin=*,nosep,topsep=0pt]
  \item Implements and attachments
  \item Hidden, safety-critical failures
  \item Smaller farms far from service
\end{itemize}
\vspace{4pt}
\end{minipage}
&
\cellcolor{quadblue}
\begin{minipage}[t]{\linewidth}\vspace{4pt}
\centering
\textbf{High urgency + high coverage}\\[2pt]
(\textbf{Strong but constrained})\\[4pt]
\begin{itemize}[leftmargin=*,nosep,topsep=0pt]
  \item Continuous-use dairy
  \item Low false-alarm tolerance
  \item Must fit existing workflows
\end{itemize}
\vspace{4pt}
\end{minipage}
\\
\hline
\cellcolor{quadyellow}
\begin{minipage}[t]{\linewidth}\vspace{4pt}
\centering
\textbf{Low urgency + low coverage}\\[2pt]
(\textbf{Opportunistic})\\[4pt]
\begin{itemize}[leftmargin=*,nosep,topsep=0pt]
  \item Older equipment
  \item Self-repair and independent shops
  \item Minimal OEM diagnostics access
\end{itemize}
\vspace{4pt}
\end{minipage}
&
\cellcolor{quadgray}
\begin{minipage}[t]{\linewidth}\vspace{4pt}
\centering
\textbf{Low urgency + high coverage}\\[2pt]
(\textbf{Hardest to enter})\\[4pt]
\begin{itemize}[leftmargin=*,nosep,topsep=0pt]
  \item New equipment with mature OEM programs
  \item Lower incremental value for add-ons
\end{itemize}
\vspace{4pt}
\end{minipage}
\\
\hline
\end{tabular}
\caption{Qualitative segment map. Adoption potential varies with downtime urgency and existing sensing or service coverage.}
\label{fig:segment_map}
\end{figure}

\textbf{OEM ecosystems define the ceiling for retrofit value.} Manufacturer interviews described brand-specific diagnostic stacks in which fault codes, service schedules, and repair workflows are mediated entirely through dealer tooling and OEM applications. Where these ecosystems are mature, the incremental value of an external acoustic layer is small: the asset is already instrumented, the service workflow is established, and adding a third-party application creates friction without proportionate benefit. Implements and attachments, by contrast, sit outside this coverage even when the towing tractor is fully instrumented. Operators facing proprietary access restrictions also benefit from a non-invasive monitoring layer that generates evidence for triage and scheduling, even though such a layer cannot clear codes or authorize repairs.

\textbf{Operator scale determines exposure.} Small-to-medium operations face a compounding disadvantage: older or mixed fleets with less redundancy, limited in-house maintenance capacity, and greater exposure to geographic service delays. Large farms can often absorb unplanned downtime through internal staff and backup equipment; smaller farms cannot. Monitoring is therefore disproportionately valuable to the segment least served by existing diagnostic ecosystems, because it can create lead time before a minor change becomes an urgent stoppage. Continuous-use dairy contexts represent a separate high-value case: equipment running 12 or more hours per day leaves little margin for unplanned stops, and the cost of a missed failure is immediate and operational rather than seasonal. Figure~\ref{fig:operator_segment} presents this qualitative fit assessment.

Taken together, these findings motivate the segment map in Figure~\ref{fig:segment_map}. The strongest early wedge is high-urgency, low-coverage contexts: implements and attachments, hidden rotating components, and smaller operations far from service. The hardest segment to enter is low-urgency, high-coverage contexts: newer equipment with mature OEM programs, where the incremental value of an external acoustic layer is lowest. Audio-based monitoring is therefore best framed as a sensing-gap layer that operates alongside established diagnostic infrastructure rather than in competition with it.

\subsection{Low-friction data collection is essential for deployment}
Stakeholders repeatedly emphasized that participation depends on simplicity and clear benefit. A domain expert with experience in field data collection noted that pairing a student with farmers can be an effective hybrid strategy for collecting and annotating data in context, reducing burden on operators while ensuring usable labels.

\begin{quote}
\small ``A summer student, in hybrid with farmers, is the best way.'' (I8, domain expert)
\end{quote}

Other notes reinforce this theme: farmers indicated willingness to cooperate if the process is easy and clearly beneficial, for example through improved uptime or avoided repair costs. Off-season periods were described as times for deferred maintenance such as repairs, blade replacement, and oil changes, suggesting a natural window for planned data collection. Some niche failure mechanisms, such as battery wear caused by short moves that do not recharge, further indicate that seemingly minor workflow behaviors can drive recurring failures. Monitoring systems may therefore need to account for these contextual triggers.

\section{Discussion}
The preceding results identify where a retrofit acoustic layer is most likely to matter. This section turns from segment fit to system design: what must be built so that field audio becomes usable maintenance intelligence under sparse labels, variable operating conditions, and constrained repair workflows?

\subsection{Derived requirements for an adoption-grounded audio pipeline}
Operators want earlier notice of abnormal change, but the evidence suggests they are unlikely to adopt a tool that adds annotation burden, creates frequent interruptions, or produces outputs that cannot inform repair decisions. The requirements below therefore span the full pipeline, from capture and annotation to alerting, evidence generation, and feedback. Figure~\ref{fig:process} depicts the proposed pipeline that these requirements shape.

\begin{enumerate}

  \item \textbf{Low-friction, commodity-device capture.}
  Nine of ten survey respondents had never used a diagnostic or predictive tool, and interview participants consistently cited workflow burden as a barrier to adoption. The sensing front end must therefore require no specialized hardware, minimal calibration, and recordings short enough to embed in normal operational pauses. Phones and compact microphones match the deployment setting better than laboratory-grade equipment.

  \item \textbf{Mandatory operating context attached to every recording.}
  Audio is ambiguous without context: the same spectral signature may indicate normal load variation on one machine type and incipient bearing failure on another. Every clip must therefore carry structured metadata, including machine type, subsystem, task, operating load state, perceived condition, and freeform operator notes, as a first-class data requirement rather than an optional annotation (Figure~\ref{fig:app}).

  \item \textbf{Asset-specific, baseline-first inference.}
  The fault-label regime in field deployment is sparse and delayed: failures are rare, confirmation arrives only after inspection or repair, and cross-machine normalization obscures equipment-level drift. The inference layer should therefore establish a per-asset acoustic baseline before attempting fault classification, and it should flag persistent deviation from that baseline as the primary detection signal. This design is consistent with how operators themselves reason: they notice that ``something sounds different'' rather than identifying a named fault mode.

  \item \textbf{Robustness to heterogeneous field conditions.}
  Agricultural environments impose sensing variability that is absent from controlled evaluations: wind, tractor engine masking, cab acoustic isolation, implement-to-implement model differences, load-dependent frequency shifts, microphone placement inconsistency, and geography-specific wear patterns noted by manufacturers. Models must be evaluated under these conditions and should not assume stationary or clean inputs.

  \item \textbf{Temporally stable, uncertainty-aware alerting.}
  Dairy and seasonal-use contexts impose asymmetric false-alarm costs: an alert that interrupts a milking cycle or a harvest pass has direct operational cost. Single-observation triggers are therefore insufficient. Alerts should be issued only after evidence accumulates across multiple recordings, uncertainty should be surfaced rather than suppressed, and the system must distinguish persistent degradation trajectories from transient acoustic noise.

  \item \textbf{Evidence artifacts for dealer and technician triage.}
  The repair constraints identified across interviews, including proprietary diagnostic access, technician scheduling delays, and geographic service gaps, mean that an alert alone is insufficient. The pipeline must produce shareable artifacts: timestamped clips, deviation trend charts, and operating-context summaries that operators can transmit to a dealer or technician to support remote triage and accelerate service scheduling, even when OEM tooling is required for the final repair.

  \item \textbf{Closed-loop human feedback for continual label acquisition.}
  Because labeled fault data are scarce at deployment time, the system must treat operator interactions, such as confirmed repairs, ignored alerts, inspection outcomes, and false-alarm reports, as structured supervision signals. Each resolved episode should feed back into the inference layer, converting field use into the labeled corpus that downstream fault classification requires. Adoption and model improvement are therefore coupled rather than sequential.

  \item \textbf{Decision-theoretic workflow integration.}
  The stakeholder evidence consistently distinguished among high-urgency continuous-use contexts (dairy), time-sensitive seasonal windows (planting and harvest), and lower-pressure opportunistic settings. A uniform alerting threshold is inappropriate across these regimes. Escalation logic should be conditioned on the expected cost of unplanned downtime relative to the cost of a false-alarm interruption, behaving conservatively during peak operational periods and more permissively during maintenance windows.

\end{enumerate}

Several of these requirements are in tension with one another in ways that implementation must resolve. Low-friction capture conflicts with mandatory context, because a heavier annotation burden reduces the likelihood that operators record at all. Stable alerting conflicts with early warning, because suppressing jitter delays detection. Conservative escalation in high-urgency contexts conflicts with sensitivity. These tradeoffs do not have universal solutions; they require calibration per deployment context, which is itself an argument for the asset-specific, workflow-conditioned design that the requirements prescribe rather than a one-size-fits-all pipeline.

\subsection{Tradeoffs and implications}

The discovery results shift the design question from fault-detection feasibility to value under real maintenance constraints. The interviews indicate that workflow burden, service-ecosystem fit, false-alarm tolerance, and evidence usability may influence adoption as much as model accuracy.

Three contextual conditions jointly determine where monitoring creates value. First, \textit{failure visibility}: if operators can already perceive a problem from inside a modern quiet cab, an external acoustic layer adds little. Second, \textit{sensing coverage}: if an OEM already instruments the asset with fault codes and planned service, the marginal value of a retrofit layer is small. Third, \textit{service access}: if repair requires dealer software regardless, monitoring matters only when it improves the service interaction, for example by documenting symptoms, timing, and operating context before a technician arrives. These three conditions jointly explain the segmentation in Figure~\ref{fig:segment_map}. Tractors often fail all three tests, whereas implements, attachments, and continuous-use dairy equipment often pass all three. The same acoustic capability may therefore be peripheral in one segment and useful in another.

\begin{figure}[H]
    \centering
    \includegraphics[width=1.0\linewidth]{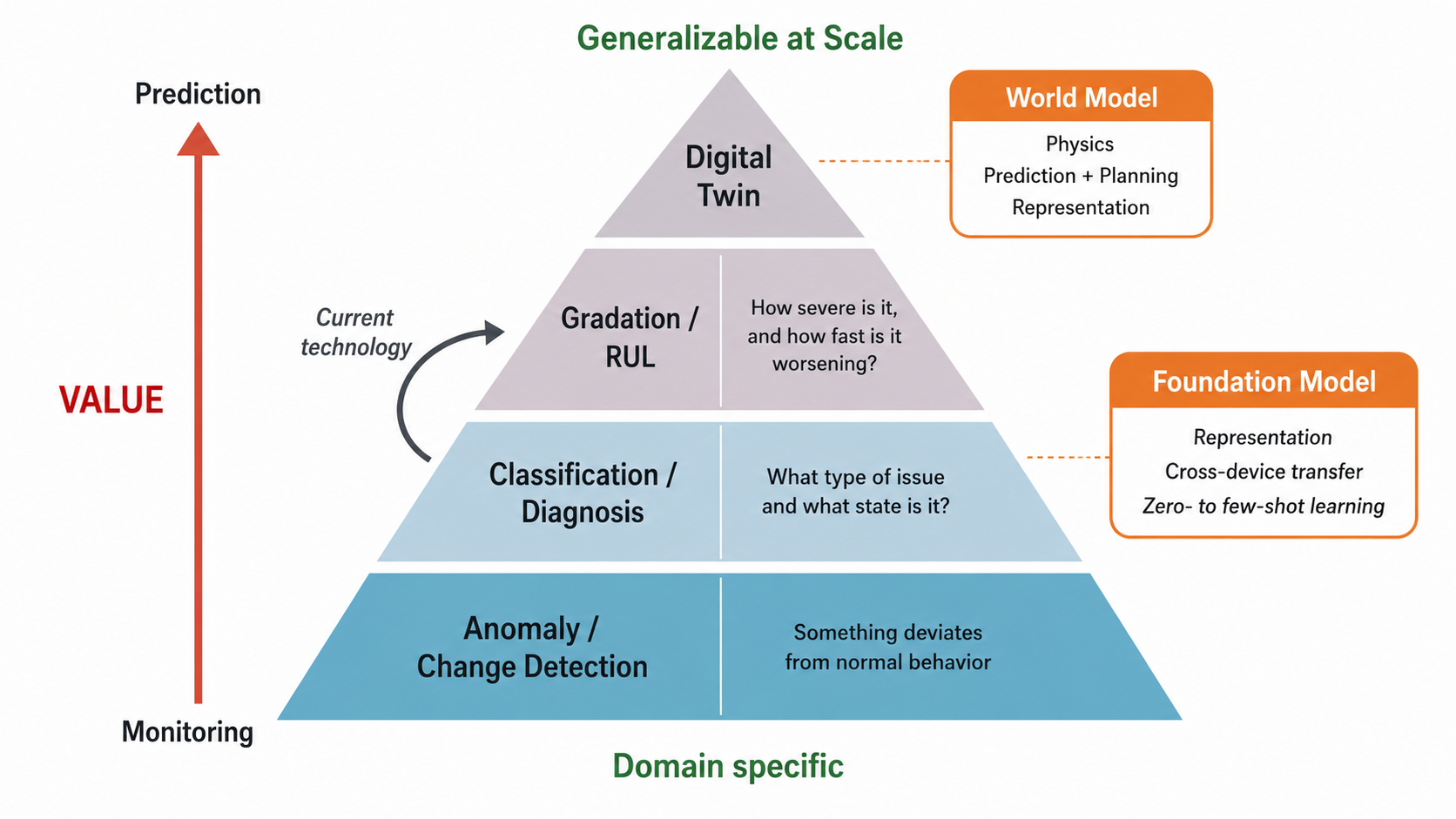}
    \caption{Conceptual hierarchy of audio-based condition monitoring capabilities for agricultural machinery. Lower tiers represent domain-specific monitoring tasks such as anomaly and change detection, while higher tiers move toward diagnosis, remaining useful life estimation, and digital-twin-based prediction. Current technologies most often support lower-to-middle capabilities, whereas foundation and world models may enable more generalizable transfer, planning, and predictive reasoning across heterogeneous equipment and operating conditions, at the cost of substantially greater data requirements.}
    \label{fig:capability_hierarchy}
\end{figure}

Given these constraints, the MVP belongs at the base of the capability hierarchy shown in Figure~\ref{fig:capability_hierarchy}: anomaly and change detection, asking only whether a machine sounds different from its normal behavior. This is the question operators already ask informally, and it is the tier that demands the least from a data and label regime that field deployment cannot yet provide. Failures in agricultural settings are rare, confirmation is delayed, and equipment heterogeneity, including model differences, geography-specific wear, and operator-driven misuse, means that higher-tier capabilities such as fault classification, gradation, and remaining useful life estimation require labeled corpora and cross-device generalization that do not yet exist for retrofit sensing. The hierarchy does not prescribe immediate development of every tier. Instead, each tier creates the data and capabilities needed by the next: base-tier anomaly detection generates the contextually labeled field data that classification and prognosis eventually require, and the closed-loop feedback requirement derived from the stakeholder evidence is the mechanism that makes that accumulation possible over time.

\begin{figure}[H]
    \centering
    \includegraphics[width=0.6\linewidth]{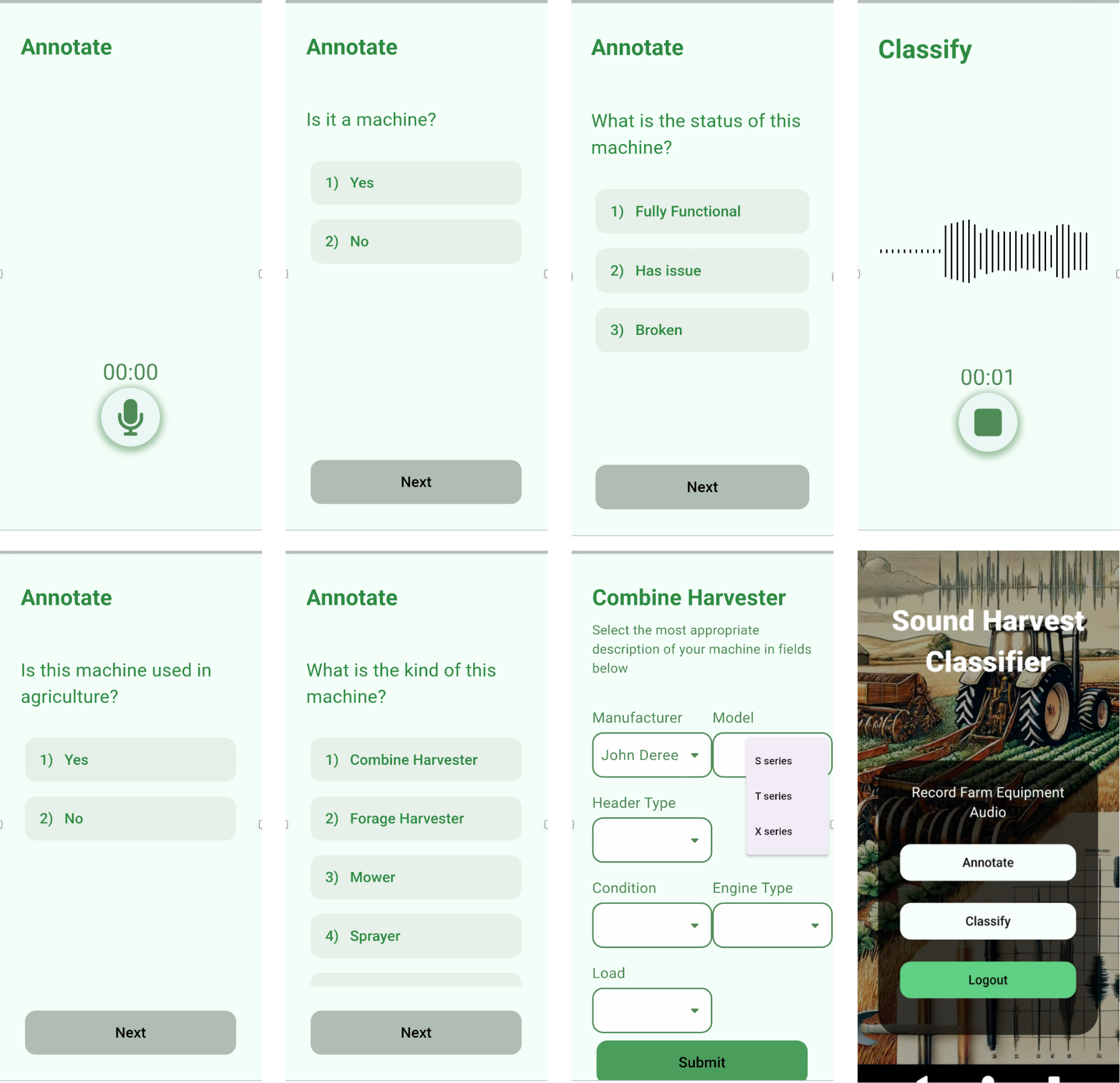}
\caption{The Sound Harvest Classifier, a mobile application developed as part of this work for in-field audio capture and annotation. Operators record short clips and attach structured labels, including machine type, subsystem, and operating condition, at the moment of capture, converting informal operator observations into a structured signal for baseline comparison, deviation tracking, and dealer triage.}
\label{fig:app}
\end{figure}

The near-term output should be a structured case record: the clip, deviation trend, timestamp, and operating context that explain why the system surfaced a change. The record helps the operator decide whether to continue, inspect, or schedule service, and it can accompany a dealer interaction even when OEM tooling is required for the final repair. The Sound Harvest Classifier (Figure~\ref{fig:app}) operationalizes the capture end of this workflow, and the proposed pipeline (Figure~\ref{fig:process}) closes the loop: field audio becomes structured signal, structured signal becomes stable alerts and case records, and confirmed outcomes become training labels for the next iteration. Adoption and model improvement are coupled by design rather than sequenced.

Technical evaluations of audio-based fault detection typically optimize detection accuracy under controlled conditions. Field deployment additionally depends on whether the system fits existing workflows, whether its outputs remain useful under proprietary service constraints, and whether use of the system produces labels for later improvement. The customer discovery results identify these operational requirements for the technical pipeline.

\section{Limitations}
This study is exploratory, and its findings should be interpreted accordingly. The sample is small and not statistically representative, and interviewees span roles with different incentives (operators, dealers, and manufacturers), which may shape the concerns each group emphasized. The survey includes both farmers and researchers, which may bias reported priorities toward those of respondents more familiar with emerging technology. The quantitative results are therefore best read as indicative rather than generalizable estimates. Nevertheless, similar themes appeared independently in the survey and the interviews, and this cross-source convergence supports the segment selection and MVP requirements as working hypotheses for subsequent deployment studies.

\section{Conclusion}
The results support a targeted retrofit approach to audio-based condition monitoring whose value depends on asset type, downtime exposure, and repair workflow. The initial segments indicated by the stakeholder evidence are implements and attachments, hidden rotating components, small-to-medium operations with limited service capacity, and continuous-use environments such as dairy. In these settings, downtime is costly, operators often cannot hear developing faults, and existing sensing coverage is incomplete.

The proposed application and CPS pipeline collect short, contextualized audio clips in the field, compare them with asset-specific normal operation, return stable deviation records, and use confirmed outcomes to improve the model. The development priorities that follow from this study are data-efficient capture, context-rich labels, baseline-first anomaly detection, robust field performance, and human-in-the-loop learning. The study identifies where acoustic monitoring may create useful lead time, complement OEM diagnostics, and generate the data required for later system development.

\section{Acknowledgment}
This work was supported by the Michigan Translational Research and Commercialization (MTRAC) Program, funded by the Michigan Strategic Fund and administered by the Michigan Economic Development Corporation (MEDC). This includes support by Grant Case-48166 of the 21st Century Jobs Trust Fund received through the MSF from the State of Michigan. This report was further supported using Federal funds under award 06-79-06452 (subaward agreement 006) from the U.S. Economic Development Administration, U.S. Department of Commerce. The statements, findings, conclusions, and recommendations are those of the author(s) and do not necessarily reflect the views of the U.S. Economic Development Administration or the U.S. Department of Commerce.
\bibliographystyle{unsrtnat}
\bibliography{references}

\end{document}